\documentclass[prb,aps,amsmath,amssymb,floatfix,superscriptaddress,longbibliography]{revtex4-2}
\usepackage{subcaption}
\usepackage{caption}
\usepackage{soul,color}
\usepackage{graphicx}
\usepackage{bm}
\newcommand{\bea}{\begin{eqnarray}}
\newcommand{\eea}{\end{eqnarray}}
\usepackage[colorlinks=true, citecolor=blue, linkcolor=blue, urlcolor=blue]{hyperref}
\usepackage[utf8]{inputenc}
\usepackage[T1]{fontenc}
\usepackage{mathptmx}

\begin{document}


\title[]{ Four-body singlet potential energy surface for reactions of calcium monofluoride }

\author{Dibyendu Sardar}
\affiliation{JILA, University of Colorado, Boulder, Colorado 80309, USA}
\author{Arthur Christianen}
\affiliation{Max-Planck-Institut f\"{u}r Quantenoptik, Hans-Kopfermann-Strasse 1, D-85748 Garching, Germany.}
\affiliation{Munich Center for Quantum Science and Technology (MCQST), Schellingstraße 4, D-80799 Munich, Germany}
\author{Hui Li} 
\affiliation{JILA, University of Colorado, Boulder, Colorado 80309, USA}  
\author{John L. Bohn} \email{bohn@murphy.colorado.edu}
\affiliation{JILA, University of Colorado, Boulder, Colorado 80309, USA}

\date{\today}

\begin{abstract}
A full six-dimensional Born-Oppenheimer singlet potential energy surface is constructed for the reaction CaF + CaF $\rightarrow$ CaF$_2$ + Ca using a multireference configuration interaction (MRCI) electronic structure calculation.  The {\it ab initio} data thus calculated are interpolated by Gaussian process (GP) regression. The four-body potential energy surface features one $D_{2h}$ global minimum and one $C_s$ local minimum, connected by a barrierless transition state that lends insight to the reaction mechanism.   This surface is intended to be of use in understanding ultracold chemistry of CaF molecules.

\end{abstract}

\maketitle
\section{Introduction}
A recurring goal in the science of ultracold molecules is the possibility to control their chemical reactions, by exploiting control over both internal and motional degrees of freedom.  The alkali dimer KRb was an early workhorse in this effort. Its reaction rates have been controlled by temperature \cite{ospelkaus2010, de2019degenerate}, quantum statistics \cite{valtolina2020dipolar}, electric field \cite{ni2010dipolar, quemener2010strong,quemener2011dyn, wang2015tuning, matsuda2020resonant} and optical lattices \cite{de2011controlling, tobias2022reactions}.  Moreover, a complete survey of the reaction products K$_2$ and Rb$_2$, including their branching ratios, has been performed \cite{hu2019direct, hu2021nuclear, liu2021precision}.  The lifetimes of the collision complex that separates reactants and products in the KRb system have also been measured \cite{liu2020photo}.  

Beyond this, a host of experiments involving ultracold alkali dimers have observed anomalously long lifetime of the collision complex, whose origins remain elusive \cite{gregory2020loss, bause2021collisions, gersema2021probing}.  Including these experiments under the general umbrella of ultracold chemistry, this chemistry can be controlled by means of electric field \cite{quemener2016shielding, gonzalez2017adimensional} or microwave \cite{karman2018microwave} shielding, preventing the molecules from ever getting close enough to react.  Such methods have been demonstrated in several labs \cite{matsuda2020resonant, anderegg2021observation, li2021tuning, schindewolf2022evaporation}. 

The arsenal of molecules that will be available for control of this type is growing rapidly, owing to new advances in laser cooling that allow one to cool and trap even large polyatomic species \cite{shuman2010laser, kozyryev2017sisyphus, anderegg2018laser, cheuk2018lambda, mitra2020direct, ding2020sub, augenbraun2020molecular, baum20201d, vilas2022magneto}.  Among these laser-cooled species, we focus here on the CaF molecule, which has been cooled and trapped in optical tweezers \cite{zhelyazkova2014laser, truppe2017molecules, anderegg2019optical}, raising yet another possibility, that of merging the tweezers and instigating chemical reactions on demand \cite{cheuk2020observation}.  These molecules are subject to the exothermic reaction
\begin{equation}
\mathrm{CaF} + \mathrm{CaF} \rightarrow \mathrm{CaF}_2 + \mathrm{F},
\end{equation}
releasing approximately $4300$ cm$^{-1}$ of energy.  The product CaF$_2$ is widely used in optical applications such as window glass and lenses, although this is of no consequence to the present discussion.

Significantly, this reaction is only possible in the singlet electronic  state of the Ca$_2$F$_2$ tetramer; the excited triplet state is immune to the reaction.  This raises the possibility of controlling the reaction by careful state preparation of the initial spin states of the reactants, a possibility that has been considered in various contexts \cite{sikorsky2018spin, mohammadi2021life, hu2021nuclear, son2022control}.   This kind of control could present refined opportunities to, for example, alter branching ratios to the various ro-vibrational states of the product. 

Conceptually, evaluating ultracold chemical reactions splits into considerations of long-range and short-range physics.  In the long range, typified by reactants further apart than their van der Waals length, the molecules move slowly and their interactions are given by comparatively weak and well-characterized long-range forces.  In this limit, scattering wave functions are comparatively easily calculated.  It may be said that propagation from infinity, through this long-range region, can be exploited to prepare the molecules for their entrance into the short-range, chemical cauldron where the reaction takes place.  In this short-range region, forces among the atoms are too strong to be under experimental control, and the native dynamics takes over.  It  is therefore important to know something of the short-range dynamics, so that one may try to  plan the long range control that sets the molecules on a desired path.

This article takes the first step in this endeavor, by computing a complete {\it ab initio} short-range singlet potential energy surface for the CaF-CaF reaction, to help us better understand the energetics and reaction pathways we may attempt to exploit in the name of control.    We note that in the similar context of SrF-SrF cold collisions, a reduced surface was proposed some years ago \cite{meyer2011chemical}. This reference identified a ``handoff''  mechanism, whereby one of the F atoms could find itself in a double well potential in the no-man's-land between the alkaline earth atoms; from here the F could go either way, either completing the reaction or remaining with its initial partner.  The calculation of Ref.~\cite{meyer2011chemical} was limited to a one-dimensional investigation, however.

In this article, we construct  a full six-dimensional singlet potential energy surface (PES) for the Ca$_2$F$_2$ dimer.  We identify global and local minima in this PES, as well as a reaction path that takes the system between them via a submerged barrier.  Based on the minimum-energy pathway, we propose a plausible reaction path through which the reaction proceeds by a torsional movement, presenting a richer dynamics than the handoff mechanism of Ref.~\cite{meyer2011chemical}.   A key feature that makes the characterization of this complicated surface possible is that the {\it ab inito} points, obtained from a MOLPRO calculation, can be readily interpolated using Gaussian Process (GP) interpolation, a technique pioneered in this context by Krems and his co-workers \cite{arthur2019,gauss-51d,gauss-dai}.

The presentation of the paper is as follows. In the section \ref{sec:level1}, we describe the \textit{ab initio} calculation on the diatomic, triatomic and tetratomic molecular systems. In section \ref{sec:level2}, we discuss some fundamental features of the GP interpolation including the symmetrization criteria of the Ca$_2$F$_2$ surface. The construction of the minimum-energy path and a plausible reaction mechanism of the Ca$_2$F$_2$ surface are discussed in the section \ref{sec:level3}. Finally, we make some conclusions in section \ref{sec:level4}.

\section{\label{sec:level1}Ab Initio Calculations}

In this section, we will describe the \textit{ ab initio} calculations of CaF-CaF from the reactant side and CaF$_2$-Ca from the product side for the concerned reaction: CaF + CaF $\rightarrow$ CaF$_2$ + Ca.

For each arrangement of nuclei in a grid to be described below, an electronic structure calculation is performed using the  MOLPRO 2012.1 software package \cite{werner2008} . The electronic configuration of ground-state Ca and F atoms are expressed as [Ar]4s$^2$ and [He]2s$^2$2p$^5$, respectively.   Thus for the \textit{ ab initio} calculations we perform a complete active space self-consistent field calculation (CASSCF) comprising the atomic orbitals 4s of Ca and 2p of F in the active space. The 3s and 3p of Ca, as well as the 1s and 2s of F, were frozen at the Hartree-Fock (HF) level of theory. Thereafter, we carry out an internally contracted multireference configuration interaction (MRCI) step where single and double excitations are taken relative to this CASSCF reference function where only the 1s electrons of F were not correlated.  The MRCI calculation incorporates  an additional Davidson correction that approximately accounts for the size consistency and higher excitations. 

The basis set for Ca is Peterson's pseudopotential-based correlation consistent polarized weighted core valence triple-$\zeta$ basis set (cc-pwCVTZ-PP) \cite{kirk2017}, where the inner core electrons are described by the Stuttgart / Koeln effective core potential ([ECP10MDF]) \cite{lim2005}. For the F atom, we consider correlation consistent polarized valence triple basis set (aug-cc-PVTZ) \cite{kendall1992} with diffuse augmenting functions.   While the method just described is used in constructing the final potential energy surface, nevertheless at various stages we vary the approach to test its adequacy, as we will see in the following subsections.

\subsection{Diatomic and triatomic molecules}

To assess the adequacy of the basis set and method, we first apply them to the three possible diatomic molecules, as well as the product CaF$_2$.
In table.\ref{tb1} we make a comparative study of the CaF, Ca$_2$, and F$_2$ molecules in terms of the equilibrium bond length ($r_e$) and depth of the well ($D_e$), comparing these with previous theoretical and experimental values. In our current method of calculation, the bond length for each of the diatomic molecules  agree with that of literature reported theoretical and experimental value with an error less than 1\% which, following common practice in the field, we accept as adequate. With regard to the well depth $D_e$, our calculations differ from other results by amounts less than $\sim 1000$ cm$^{-1}$, We therefore take this as an estimate of the uncertainty of the calculations.  Note that the largest uncertainty belongs to the F$_2$ molecule.

\begin{table}[h!]
\centering
\caption{The comparison of optimized diatom bond lengths $r_e$ in Bohr and depth of the wells $D_e$ in cm$^{-1}$ with previous theoretical and experimental data. }
\begin{tabular}{|c| c| c| c| c| c|} 
 \hline
  & CaF & Ca$_2$ &  F$_2$ \\ [0.5ex] 
           &\hspace{0.1cm}$r_e$\hspace{1.0cm} $D_e$& \hspace{-0.4cm}$r_e$, \hspace{0.7cm}$D_e$&\hspace{-0.4cm}$r_e$, \hspace{0.7cm}$D_e$\\
 \hline
 this work & 3.704, \hspace{0.1cm} 43672 & 8.101, \hspace{0.1cm} 1134 &2.671, \hspace{0.1cm} 12464  \\ 
 theory & 3.691 \cite{hou2018}, \hspace{0.1cm} 44203 \cite{hou2018} & 8.085 \cite{allard2003}, \hspace{0.1cm} 1102 \cite{allard2003} &2.665 \cite{meyer2011chemical}, \hspace{0.1cm} 12880 \cite{meyer2011chemical} \\ 
  expt & \hspace{-0.5cm}3.717 \cite{book}, \hspace{0.1cm}------- & 8.090 \cite{balfour1975}, \hspace{0.1cm} 1075 \cite{balfour1975} $\pm$150 &2.677 \cite{chen1985negative}, \hspace{0.1cm} 13410 \cite{jie2005}  \\ [1ex]   
 \hline
\end{tabular}
\label{tb1}
\end{table}

Similarly, we investigate the global minimum of the  potential energy surface of the CaF$_2$ product, as summarized in Table  \ref{tb2}.  This table compares the equilibrium  bond lengths and bond angle for the ground singlet state of CaF$_2$ to various values found in the literature.  As for the diatomic molecules, we find agreement of the bond lengths below the percent level.  The bond angle in CaF$_2$ remains a matter of some controversy. In various  theoretical studies on calcium difluoride \cite{seijo1991,kaupp1992-other,levy2000,von2002,kaupp1991,koput2004}, the ground state is found to be either linear or slightly bent,  depending on the quality and size of the basis sets used. The difference in energy between its linear and bent geometry is typically of order tens of cm$^{-1}$, making the ultimate determination uncertain at present.  In our own calculation, the minimum at 168\textdegree is separated from the linear geometry by a barrier of only 6 cm$^{-1}$, meaning our calculation is unlikely to resolve this controversy.

\begin{table}[h!]
\centering
\caption{Optimized molecular parameters of calcium difluoride where bond lengths ($r_1$ and $r_2$) are in Bohr and bond angle ($\beta$) is in degree. }
\begin{tabular}{ |c|c|c|c|c| } 

\hline
 &Symmetry & $r_1$ & $r_2$ & $\beta$ \\
\hline
this work &${}^1A_1$ & 3.804     & 3.804 & 168.10 \\ 
theory\cite{kaupp1991}&,, &  3.799    & 3.799  & 157.50\\ 
theory\cite{koput2004}&,, &  3.790    & 3.790  & 154.86\\ 

\hline
\end{tabular}
\label{tb2}
\end{table}

\subsection{Four-body surface}

Since CaF is an open shell molecule having electronic spin 1/2,  its dimer either can be a spin singlet ($S = 0$) or a spin triplet state ($S = 1$). In the present calculation, we focus exclusively on the singlet surface as it is the reactive one. 

\begin{figure}
  \includegraphics[width=0.5\textwidth]{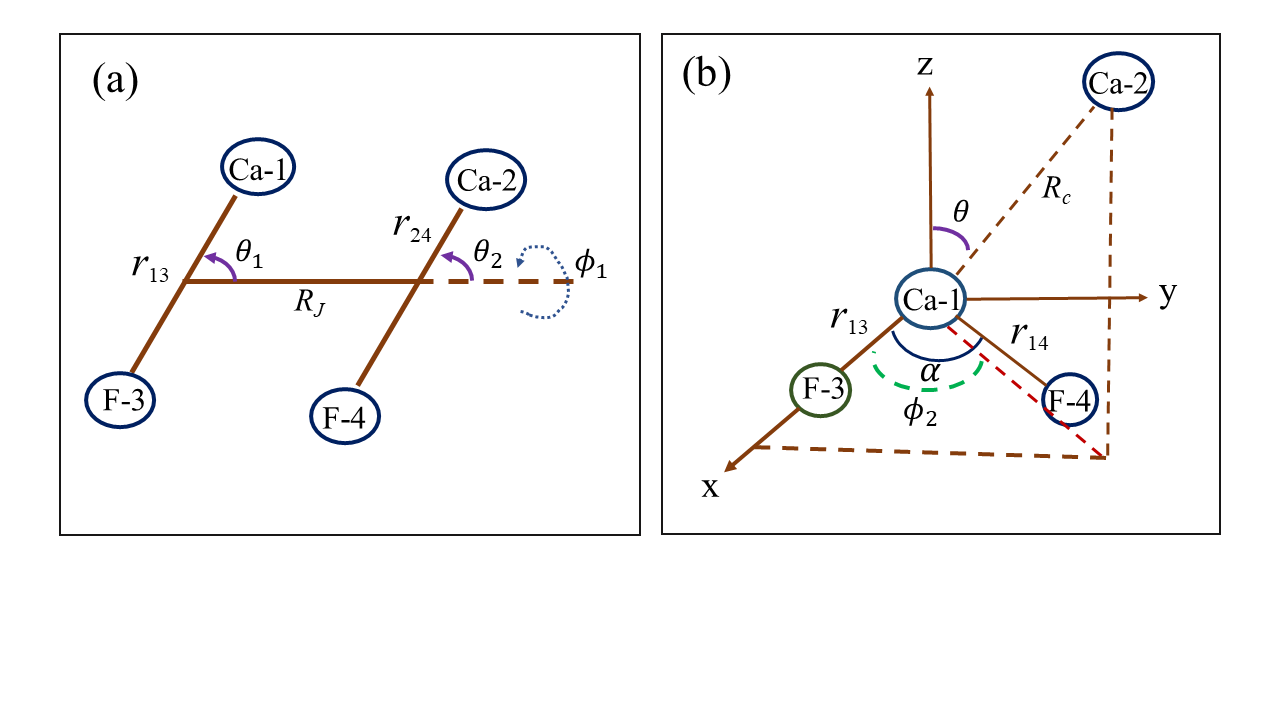}
  \caption{In the panel-(a), the Jacobi coordinates of the complex in one of the CaF-CaF arrangements are shown. The spherical polar coordinates are represented in the panel-(b) for one of the arrangements of CaF$_2$-Ca}
  \label{fig_1}
\end{figure}

The electronic structure calculations are performed on a grid of relative atoms coordinates.  For later interpolation it will be useful if these are roughly equidistant in the parameter space. For this purpose it is useful to consider the coordinate systems which describe well the two asymptotic arrangements of the reactant and product. We choose two suitable coordinate systems to describe the arrangement of atoms for the reactant and product side.  We use Jacobi coordinates for the CaF-CaF arrangement as shown in Fig.\ref{fig_1}a, where $R_{J}$ is the distance between two centers of mass (CM) of two monomer CaF molecules having bond lengths $r_{13}$ and $r_{24}$, respectively. The  molecules are tilted with respect to the intermolecular axis by  polar angles  $\theta_1$, $\theta_2$, while $\phi_1$ is the torsion angle. For the product side (CaF$_2$-Ca) we use spherical polar coodinates, as depicted in Fig.\ref{fig_1}b. Here, $R_S$ is the distance between two Ca atoms.  The bond lengths $r_1$ and $r_2$ represent the distance between the atoms Ca1-F3 and Ca1-F4. The parameter $\alpha$ is the angle between the atoms F3-Ca1-F4, $\theta$ and $\phi_2$ are the azimuthal and polar angles, respectively. 


\begin{table}[h!]
\centering
\caption{The energy of the PES Ca$_2$F$_2$ in cm$^{-1}$ for the global and local minima with optimized bond lengths (in Bohr). The energies are referred to the CaF + CaF dissociation threshold.}
\begin{tabular}{ |c|c|c|c|c| c|c|c|c|} 

\hline
method &geometry& $r_{12}$&  $r_{13}$ & $r_{34}$ & r$_{23}$  & r$_{24}$ &r$_{14}$ &E$_{min}$  \\
\hline
CCSD(T)&  D$_{2h}$& 6.360 & 4.057 & 5.039 &   -     &   -    &   -   & -18923 \\
MRCI   &          & 6.401 & 4.053 & 4.979 &   -   &   -    &   -     &-17936 \\
\hline
CCSD(T)&  C$_s$   & 6.522 & 3.756 & 7.337 &  4.076   &  4.007 & 10.268  & -15363 \\
MRCI   &          & 6.606 & 3.743 & 7.327 &  4.064  &   4.006 & 10.337   &-14924 \\
\hline
\end{tabular}
\label{tb3}
\end{table}

Initially, we calculate the optimized geometries of the four-body surface incorporating the  basis set described above. Using the method of geometry optimization in MOLPRO \cite{werner2008}, we find one global minimum and one local minimum of the four-body surface Ca$_2$F$_2$. At present, we are unaware of literature values of either global or local minima of the Ca$_2$F$_2$ surface. Therefore, to estimate the uncertainty in our current method of calculation, we determine the same optimization for the global and local minima by another level of  theory, coupled-cluster singles, doubles and perturbative triples excitation [CCSD(T)] using the same basis sets: Ca = cc-pwCVTZ-PP and F = aug-cc-PVTZ. For both of the minima, the geometric parameters and corresponding energies are tabulated in Table \ref{tb3} for both \textit{ab initio} methods.

Both  calculations confirm that the global minimum of the Ca$_2$F$_2$ surface has a D$_{2h}$ symmetry with a rhombic structure for which the bond lengths satisfy $r_{13} = r_{24} = r_{23} = r_{14}$, and that the local minimum has C$_s$ symmetry. The two calculations moreover agree on the bond lengths at the usual percent level of accuracy.

The energies of these minima, however, differ by up to $\sim 1000$ cm$^{-1}$, with the CCSD(T) calculation consistently finding deeper minima. It may be presumed that the CCSD(T) calculation is more accurate, because it includes higher order electronic correlations. However, CCSD(T) is not generally well-behaved across a reaction pathway, and therefore using CCSD(T) for the entire PES will expectedly lead to irregularities. We therefore chose to carry out the calculations on the MRCI level. 
In this context, we therefore use the results of Table \ref{tb3} to estimate the uncertainty of our calculation as on the order of $\sim 1000$ cm $^{-1}$, consistent in fact with the diatomic molecule calculations.  

Finally, we check the quality of our chosen basis set for Ca which is used throughout our \textit{ab initio} calculation.  The comparison of the global well depth in our standard  cc-pwCVTZ-PP basis set for Ca is compared to three other basis sets with longer acronyms in Table \ref{tb4}.  The basis set aug-cc-PVTZ is assumed for F throughout.  The difference so induced is seen to be of order $\sim 100$ cm$^{-1}$, whereby the influence of the Ca basis set is not regarded as a significant source of uncertainty.  
 

\begin{table}[h!]
\centering
\caption{The comparison of well depth (D$_e$) in cm$^{-1}$ with different basis sets of Ca for the global minimum of Ca$_2$F$_2$.}
\begin{tabular}{ |c|c|c|c|c| } 

\hline
Serial No.& Basis & Depth of well (D$_e$) & error (\%)\\
\hline
1 & cc-pwCVTZ-PP     & 17936 & ---\\ 
2 & cc-pCVQZ-PP \cite{kirk2017}      & 17787 & 0.83\\ 
3 & cc-pwCVQZ-PP \cite{kirk2017}    & 17885 & 0.29\\ 
4 & aug-cc-pwCVQZ-PP \cite{kirk2017} & 17927 & 0.05\\ 

\hline
\end{tabular}
\label{tb4}
\end{table}

Thus encouraged that the electronic structure calculations are sound, we proceed to use the MRCI method with the basis sets described above, to calculate the four-body potential energy surface.   Focusing here on the short-range part of the surface relevant to the chemical reaciton, we limit the range of atomic coordinates as follows:
 For the CaF-CaF reactant arrangement, we consider coordinates bounded by: 1) monomer bond lengths of CaF ($r_{13}$, $r_{24}$) vary from 3.2 a$_0$ to 7 a$_0$ with $r_{24} > r_{13}$; 2) $R_J$ varies from 4.0 a$_0$ to 18 a$_0$; the angles $\theta_1$, $\theta_2$, and $\phi_1$ vary from 0 to $\pi$.  The corresponding range of product coordinates is given by: 1)  3.3 a$_0$ $<r_{13}<$ 7 a$_0$,; 2) 3.4 a$_0$ $<r_{14}<$ 7.2 a$_0$,; 3) 4 a$_0$ $<R_S<$ 15 a$_0$;  and 4)  0 $<\alpha, \theta, \phi_2<$ $\pi$.

\section{\label{sec:level2}Machine-learning algorithm: Gaussian process regression}
Traditionally, interpolation of potential energy surfaces was an art form that required careful assessment of the appropriate fitting functions in various regions of configuration space, followed by accurate determination of the parameters of these fitting functions.  This situation has changed recently with the use of Gaussian process (GP) interpolation methods, pioneered in this context by Krems and collaborators \cite{cui2016}.  In these methods, the computed {\it ab initio} values of the PES at a set of atom coordinates allows a direct estimate of the value of the PES at a new point at which the {\it ab initio} calculation was not performed. This process proceeds directly without the intermediate step of determining a fitting function.
For the final potential energy surface in the GP fitting we use inverse atomic distance coordinates. Because, it is difficult to describe different chemical arrangements of the four-body system equivalently by Jacobi coordinates, making it troublesome to fit the reactive part of the PES \cite{arthur2019}. Therefore, we choose inverse interatomic spacing coordinates.

Let ${\bf x}$ denote a set of six coordinates that defines a configuration of atoms  (e.g., these are the inverse interatomic spacings in the coordinate system we use).    There is presumed to exist a continuous PES in these coordinates, denoted $V({\bf x})$.  The {\it ab initio} calculation provides a discrete set of such values, $y_i = V({\bf x}_i)$, referred to as the training set in the machine learning argot.  The {\it possible} values of the PES at the points ${\bf x}_i$ are assumed to satisfy some Gaussian-like probability distribution
\begin{equation}
P(\{V_i\}) \propto \exp \left[ - \frac{ 1 }{ 2 } \sum_{ij} y_i K^{-1}({\bf x}_i,{\bf x}_j) y_j \right].
\end{equation}
Here  the function $K$, called a kernel, describes the covariance of the variables in this distribution.   Significantly, the kernel is a function of the values ${\bf x}_i$ at which the data are computed, and not of the computed values $V_i$ themselves.  The functional form of $K$ is relevant to the quality of the fit.  Typically it contains parameters such as a characteristic length scale, so that, for example, the data at two points ${\bf x}_i$ and ${\bf x}_j$ are correlated only if $|{\bf x}_i - {\bf x}_j|$ is smaller than this scale,  Thus the kernel can describe features like the smoothness of the fit, but without specifying a particular fitting function. 

In the interpolation step, it is desired to know the predicted value $y^*$ of the PES at a new configuration ${\bf x}^*$ at which the {\it ab initio} calculation has not been performed.  To make this interpolation,  It is assumed that the additional point would satisfy the same distribution with a kernel of the same functional form.  In this case a standard algebraic manipulation yields a distribution for the unknown value $y^*$, with mean value and standard deviation \cite{rasmussen} 
\begin{equation}
 \mu({\bf x_*}) = K({\bf x_*},{\bf x})^T \left[K({\bf x},{\bf x}) + \sigma^2_n I \right]^{-1} \bf y 
\end{equation}
\begin{equation}
 \sigma({\bf x_*}) = K({\bf x_*},{\bf x_*}) - K({\bf x_*},{\bf x})^T\left[K({\bf x},{\bf x}+ \sigma^2_n I)\right]^{-1} K({\bf x_*},{\bf x})
\end{equation}

There are many functional forms possible for the kernel function.  Each such case, the kernel depends on some set of parameters,  which are optimized by
 maximizing the log marginal likelihood:
\begin{equation}
 \log p({\bf y}|{\bf x; \theta_3}) = -\frac{1}{2}{\bf y}^TK^{-1}{\bf y} - \frac{1}{2}\log|K| - \frac{n}{2}\log(2\pi)
\end{equation}
where $|K|$ is the determinant of K and $\theta_3$ denotes the collective set of parameters for the analytical function of $\Bbbk$ (see below).

The optimum quality of GP fit for a multidimensional surface depends on the kernel and the coordinate representation. Here, we consider inverse atomic distance coordinates for GP fitting. 
Following the GP fits of potential energy surfaces in \cite{arthur2019}, the kernel we use is the product of a Mat$\acute{e}$rn kernel \cite{rasmussen} and Constant kernel \cite{rasmussen}.  This kernel takes the form 
\begin{equation}
 \Bbbk({\bf x}_i, {\bf x}_j) = C M({\bf x}_i, {\bf x}_j,l_1, l_2, l_3, l_4, l_5, l_6)
\end{equation}
where $C$ is the Constant kernel. Here, $M$ is the anisotropic Mat$\acute{e}$rn kernel depending on the parameter ($\nu$) besides length scales \cite{rasmussen}. For $\nu = 2.5$, the Mat$\acute{e}$rn kernel is defined as \cite{rasmussen,arthur2019} 
\begin{equation}
 M({\bf x}_i, {\bf x}_j) = \sum_{{k} = 1}^{6} \left[ 1 + \sqrt{5}l_k^{-1} |x_{i,k} - x_{j,k}| + \frac{5}{3}l_k^{-2}|x_{i,k} - x_{j,k}|^2 \right] \times  exp \left( -\sqrt{5}l_k^{-2} |x_{i,k} - x_{j,k}|^2 \right)
\end{equation}
The characteristic length scales $l_1 -l_6$ are the parameter of $\theta_3$.

In general, validity of the GP model increases monotonically with an increase in the number of training points. However, for evaluating the PES with $n$ training points, GP involves an iterative inversion of a square matrix $n \times n$, scaling as $O(n^3)$. But, numerical evaluation of a GP model can be reduced to a product of two vectors of size $n$, scaling as $O(n)$. Therefore, one needs to care about $n$ so that the maximum number of training points should lie in the configuration space. Besides this, insertion of the symmetrization in the GP model is essential to explore the reactive part of the surface entirely since  
GP has no prior information about the symmetry.

\subsection{Symmetrization of GP}
The PES as interpolated by the Gaussian process must be symmetric under the exchange of either the two Ca nuclei, or the two F nuclei. Accounting for this symmetry can reduce the configuration space over which the fit must be performed. Besides, we add some symmetrically equivalent points to the training set to improve the quality around the symmetrization boundaries. Let us define the permutation operators $\hat P_{12}$ that interchange the Ca nuclei and $\hat P_{34}$ that exchange the F nuclei. Then the permutation operators $\hat P_{12}$, $\hat P_{34}$, and $\hat P_{12}\hat P_{34}$ give equivalent energy when two nuclei permute. 
For the  CaF-CaF arrangement, we consider training point region where $r_{13}<r_{24}$ and also fulfills $(r_{13} + r_{24}) < (r_{14} + r_{23})$ since $r_{13}$, $r_{24}$ are the bond lengths of each CaF monomer in the chosen Jacobi coordinate. In the spherical polar coordinates of CaF${_2}$+Ca arrangement, the training points region is considered where $r_{13}$ < $r_{14}$.

In our GP model, the symmetrization procedure is accomplished by coupling different contributions utilising a switching function. Suppose, we consider two functions $ F_1 (\vec x)$ and $F_2 (\vec x)$ between which switch is made, then
\begin{equation}
 F_m\left[u; m, w, F_1 (\vec x), F_2 (\vec x)\right] = z(u, c, w)F_1 (\vec x) + \left[1-z(u, c, w)\right]F_2 (\vec x) 
\end{equation}
where the sigmoid function $z$ should be twice differentiable and switches within the finite interval $(c-w) < u <(c+w)$, it can be expressed as \cite{arthur2019} 
\[
   z(u, c, w) =
  \Biggl\{ {\begin{array}{cc}
  \hspace{-5.2cm}   & $if$ \hspace{0.2cm} u \le c-w   \\ 
   \frac{1}{2} + \frac{9}{16}{\sin\frac{\pi(u-c)}{2w} + \frac{1}{16}\sin\frac{3\pi(u-c)}{2w} }, & $if$ \hspace{0.2cm} c-w < u < c+w  \\
   \hspace{-5.2cm}  & $if$ \hspace{0.2cm} u \ge c+w \\
  \end{array} } 
\]
Here, $u$ is the parameterizing parameter for switching, $c$ is the value of $u$ around which switch is centered and $w$ is halfwidth of the switching interval. The symmetrization scheme for CaF-CaF region is given as
\begin{equation}
 V^{GP}_1(\vec x) = F_m\left[\frac{r_{13}}{r_{13}+r_{24}}; \frac{1}{2}, \frac{1}{16},  V^{GP}(\vec x),  V^{GP}(\hat P_{12}\hat P_{34}\vec x)\right]
\end{equation}
\begin{equation}
 V^{GP}_2(\vec x) = F_m\left[\frac{r_{23}}{r_{23}+r_{14}}; \frac{1}{2}, \frac{1}{16},  V^{GP}(\hat P_{12} \vec x),  V^{GP}(\hat P_{34}\vec x)\right]
\end{equation}
\begin{equation}
 V^{GP}_{CaF-CaF}(\vec x) = F_m\left[\frac{r_{13}+r_{24}}{r_{13}+r_{24}+r_{23}+r_{14}}; \frac{1}{2}, \frac{1}{16},  V^{GP}_1(\vec x),  V^{GP}_{2}(\vec x)\right]
\end{equation}
For the CaF$_2$-Ca region,
\begin{equation}
 V^{GP}_3(\vec x) = F_m\left[\frac{r_{13}}{r_{13}+r_{14}}; \frac{1}{2}, \frac{1}{16},  V^{GP}(\vec x),  V^{GP}(\hat P_{34}\vec x)\right]
\end{equation}
\begin{equation}
 V^{GP}_4(\vec x) = F_m\left[\frac{r_{23}}{r_{23}+r_{24}}; \frac{1}{2}, \frac{1}{16},  V^{GP}(\hat P_{12}\vec x),  V^{GP}(\hat P_{12}\hat P_{34}\vec x)\right]
\end{equation}
\begin{equation}
 V^{GP}_{CaF_{2}-Ca}(\vec x) = F_m\left[\frac{r_{13}+r_{14}}{r_{13}+r_{14}+r_{23}+r_{24}}; \frac{1}{2}, \frac{1}{16},  V^{GP}_3(\vec x),  V^{GP}_{4}(\vec x)\right]
\end{equation}
Finally we merge the above two arrangements as 
\begin{equation}
 V^{GP}_{12}(\vec x) = F_m\left[\frac{r_{24}}{r_{14}+r_{24}}; \frac{1}{2}, \frac{1}{16},  V^{GP}_1(\vec x),  V^{GP}_3(\vec x)\right]
\end{equation}
\begin{equation}
 V^{GP}_{34}(\vec x) = F_m\left[\frac{r_{14}}{r_{14}+r_{24}}; \frac{1}{2}, \frac{1}{16},  V^{GP}_2(\vec x),  V^{GP}_4(\vec x)\right]
\end{equation}
\begin{equation}
 V^{GP}_{total}(\vec x) = F_m\left[\frac{r_{13}}{r_{13}+r_{23}}; \frac{1}{2}, \frac{1}{16},  V^{GP}_{12}(\vec x),  V^{GP}_{34}(\vec x)\right]
\end{equation}

\begin{table}[h!]
\centering
\caption{The energy of the PES Ca$_2$F$_2$ in cm$^{-1}$ for the global and local minima with optimized bond lengths (in Bohr). }
\begin{tabular}{ |c|c|c|c|c| c|c|c|c|} 

\hline
method &geometry& $r_{12}$&  $r_{13}$ & $r_{34}$ & r$_{23}$  & r$_{24}$ &r$_{14}$ &E$_{min}$  \\
\hline
GP-I   & D$_{2h}$         & 6.355 & 4.085 & 5.136 &   -     &   -    &   -   & -18417 \\
GP-II  &                 & 6.400 & 4.043 & 4.958 &   -     &   -    &   -   & -17847 \\
\hline
GP-I   & C$_s$    & 6.320 & 3.678 & 7.046 &  4.176   &  4.140 &  9.942  & -14907 \\
GP-II  &          & 6.323 & 3.648 & 6.994 &  4.336   &  4.058 &  9.900  & -14330 \\

\hline
\end{tabular}
\label{tb5}
\end{table}

\subsection{Interpolation of surface by GP models}
For the construction of the GP model, the \textit{ab initio} energies are calculated on  selected grids, respecting the symmetrization as described above.  The grid points are chosen within the range of Jacobi and spherical coordinates as described above, and based on a  sampling method developed in Ref \cite{arthur2019}. 
In this method, we initially perform the HF calculations on a random  but roughly equidistant grid using the method of latin hypercube sampling \cite{stein1987}).  We reject grid points where the energy of the PES is above a cutoff energy, as being both irrelevant to the surface, and problematic to interpolate.  The selection of the cutoff energy in different regions in the parameter space is considered in the following way: If any of the interparticle distances between Ca and F is smaller than 3.4 $a_0$ and if R<10 $a_0$, then we use a cutoff of 5000 cm$^{-1}$. If one of these two criteria is false we use a lower value of the cutoff, given by 1000 cm$^{-1}$. This procedure is chosen so that a proper repulsive barrier is included in the short range without wasting training points to describe the vibrational potential of the free diatoms in the long range.
At the points that remain, we construct the training set using the full MRCI algorithm described above using inverse atomic distance coordinate. 



To test the accuracy of the GP interpolation, we optimize the global and local minima  of the PES Ca$_2$F$_2$ for two distinct sets of training points, referring to the resulting fits as GP-I and GP-II.  These two sets consist of  2596 and 3061 training points, respectively. We use the steepest gradient descent approach for optimization. As reported in Table \ref{tb5},  the optimized energies of the two minima agree within  $\sim 500$ cm$^{-1}$ with the {\it ab inito} calculation-, for both GP fits. The error is comparable to the uncertainty in the electronic structure calculations.  This suggests that the number of training points is adequate.

\begin{figure}
  \includegraphics[width=\linewidth]{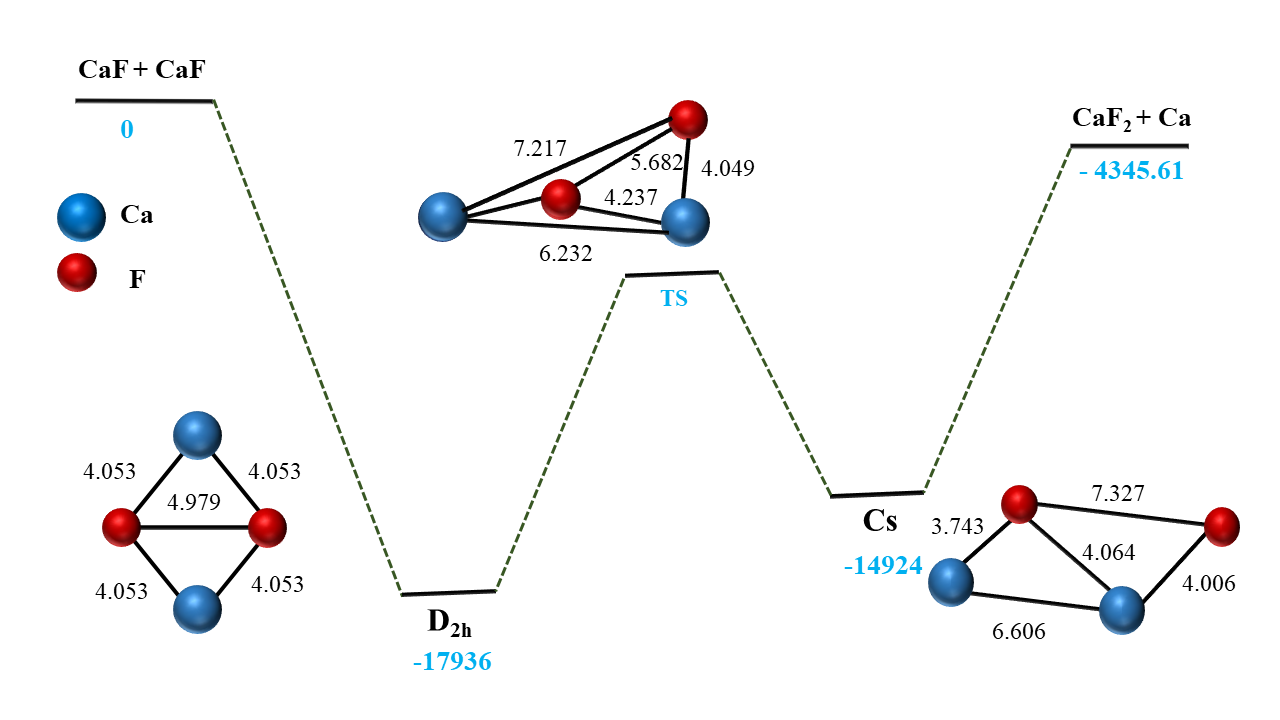}
  \caption{A reaction pathway for the reaction CaF + CaF $\rightarrow$ CaF$_2$ + Ca with energies and geometries of the stationary points. The red and blue ball indicates the atoms Ca and F, respectively. The distance between two atoms is in Bohr. The energies are in cm$^{-1}$ with respect to the CaF-CaF threshold. }
  \label{rp}
\end{figure}

As a second check, an additional set of  test points is extracted from the {\it ab initio} electronic structure calculation.  We can then compare these directly to the values approximated from the GP fit, and determine the root-mean-squared error in the fit of the test points.  This error is on the order of 1150 cm$^{-1}$ for 1315 randomly chosen test point, another confirmation of the accuracy of the fit.


Based on the result including the \textit{ab initio} method of calculation, we find uncertainties on the order of $\sim 1000$ cm$^{-1}$ for both the electronic structure calculation, and the GP fit.  Therefore, assuming these uncertainties are independent, we estimate the net uncertainty to be perhaps $\sim 1400$ cm$^{-1}$, or something like 8\% of the depth of the potential.  One hopes that this provides a reasonable qualitative picture of the potential energy surface.

\section{\label{sec:level3}Minimum Energy Path}
\begin{figure}
  \includegraphics[width=0.7\textwidth]{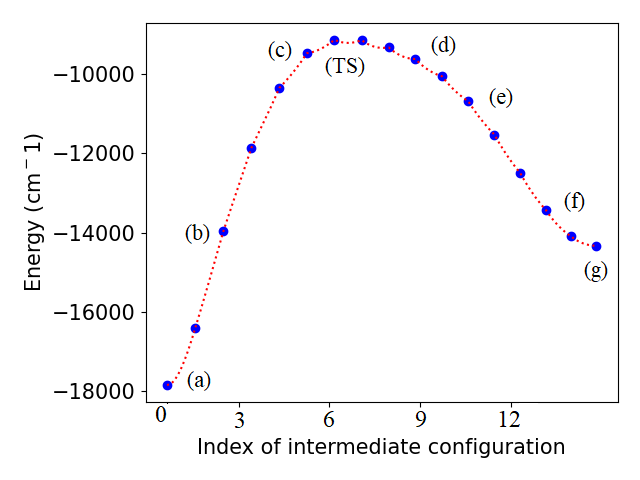}
  \caption{A transition path between D$_{2h}$ and C$_s$ minima of the Ca$_2$F$_2$ potential surface. The letters a-g represent the evolution of the geometries along the minimum-energy path.}
  \label{ts}
\end{figure}

The complete potential energy surface is therefore ready for dynamical calculations to begin, a task we defer to a future work.  However, an overview of the surface is warranted here, as a first hint of the reaction mechanism.  To this end we sketch the reaction path shown in Figure \ref{rp}.  The first thing to note is that the reaction CaF+CaF$\rightarrow$ CaF$_2$ + F is exothermic, releasing 4300 cm$^{-1}$ of energy.  

To get from reactants to products, we trace a hypothetical minimum-energy path, described by the way stations illustrated in  Figure.\ref{ts}. Some of the evolving geometries along this path are indicated by letters `a-g' and are being analysed later.  This diagram includes the global minimum of D$_{2h}$ symmetry and the local minimum of C$_s$ symmetry, as identified above. In between there exists a saddle point, with geometry and energy as shown, and which we identify in this context as a transition state.  Significantly, the energy of this state is well below the reactant and product energies, whereby the transition state occurs via a submerged barrier.  

The transition state is identified using the  ``climbing image Nudged Elastic Band (NEB)'' \cite{henkelman2000} approach.  In this method, a set of intermediate states (15 in this case) are posited between the two minima.  These points are treated conceptually as point masses, called ``beads,''subject to the energy constraint of the PES itself, and to imaginary spring forces artificially applied between each bead and the next.  The minimum-energy path between the potential minima is then obtained by minimizing the energy of this system of masses. This method produces the energy diagram in Figure \ref{ts}, constructed using the GP-II interpolation described above.  This figure shows how the energy rises and falls in going between the D$_{2h}$ minimum on the left, and the C$_s$ minimum on the right.  The blue spheres represent the locations of the beads along this trajectory, and the energies at the corresponding points of the PES.  The highest energy state along this line is therefore associated with the transition state.  The red curve is an interpolation to guide the eye. 

\begin{figure}
  \includegraphics[width=\linewidth]{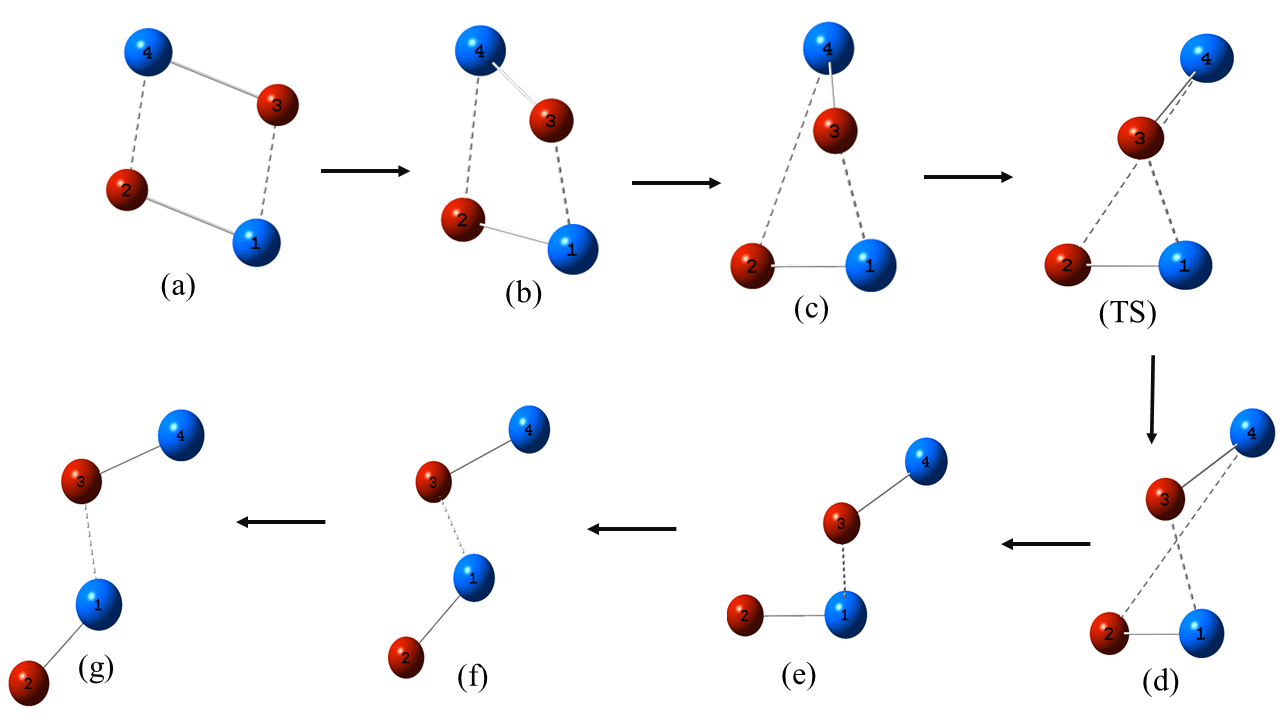}
  \caption{The evolution of the geometries on passing from global minimum (a) to local minimum (g) along the minimum-energy path is shown involving the transition state (TS)  Blue spherees represent Ca atoms, while red spheres represent F atoms. }
  \label{reaction-mechanism}
\end{figure}

 
Figure \ref{reaction-mechanism} illustrates the atomic configurations at various steps along this trajectory.  One presumes these are given in order as the reaction coordinate grows from small to large; one of them is the transition state. The geometric parameters of the TS are: $r_{12} = 6.232$, $r_{13} = 3.866$, $r_{23} = 4.237$, $r_{14} = 7.217$, $r_{24} = 4.049$, $r_{34} = 5.682$, in Bohr.  We note that the geometry of the transition state is not linear. In these drawing the blue circles represent Ca atoms, the red circles represent F atoms.  

The main result shows the evolution in geometry is to move the atoms from the D$_{2h}$ symmetry in a), to the zig-zag formation in (g). From this point, the lower calcium atom has two fluorine neighbors, which it is ready to take away as the CaF$_2$ product.  To get between these arrangements, the F atom in the upper right of (a) swings around, out of the plane of the figure, thereby moving from right to left.  Thus, at least along the reaction path shown, the reaction proceeds by a torsional movement, suggesting that the product may be put into significant rotation.  Also since the lower F atom remains more or less a spectator to the process, it might be expected that its CaF bond receives less energy than the other bond, leading to asymmetric stretch modes of the product. 

These conclusions remain, of course, speculative at this point, and will have to be assessed in dynamical calculations,  In particular, following these configurations along an adiabatic path such as this one completely disregards the influence of the energy the reactants actually have, as they commence the reaction approximately  $18,000$ cm$^{-1}$ above the global minimum.  Nevertheless, even in this approximation, it can be seen that a nontrivial re-arrangement may occur, leading to interesting ro-vibrational branching ratios, which may ultimately be the subject of control in the ultracold environment.

\section{\label{sec:level4}Conclusion}
In summary, we have constructed a six-dimensional singlet potential energy surface of the Ca$_2$F$_2$ system at short range and analyzed the features and reactivity of the surface. We observed the reaction CaF + CaF $\rightarrow$ CaF$_2$ +Ca is exothermic in nature by liberating energy $\sim$ 4300 cm$^{-1}$.  The complete construction of the Ca$_2$F$_2$ surface depends on two steps. Firstly the \textit{ab initio} points were calculated using the multireference configuration interaction method in MOLPRO and then these points were readily interpolated by GP fitting. We detected one global minimum and one local minimum having D$_{2h}$ and C$_s$ symmetry, respectively, and we constructed a minimum-energy path between these two minima that connects them via a submerged barrier. We proposed a plausible reaction pathway for the formation of the products. However, one can perform a dynamic calculation to establish the true reaction mechanism for the singlet reactive surface CaF-CaF which is beyond our scope of the study. We hope this PES would provide an initial platform to study the dynamics of the reaction. 

In future work, we will explore both the triplet surface and spin-orbit coupling to the reactive singlet surface. The triplet surface is non-reactive and suppresses the chemical reaction. Therefore, if we presume a model where initially the sample is prepared in the spin-polarized triplet state and thereafter switches on the spin-orbit coupling to the singlet surface. We anticipate the reactivity of the singlet surface where the coupling parameter is significant between these two surfaces, resulting in a chemical reaction. Therefore, the conversion of the spin-polarized non reactive surface to the non-spin-polarized reactive state can be studied including the branching ratios. This could impart a new direction in the domain of controlled chemical reactions.   


\begin{acknowledgments}
This material is based upon work supported by the National Science Foundation under Grant No. PHY 1734006.  We also acknowledge funding from an AFOSR-MURI grant as grant number GG016303..
Dibyendu Sardar especially acknowledges discussions with Kirk A. Peterson and Tijs Karman for \textit{ab initio} calculations, and we are thankful to Roman V. Krems, Jacek K\l{}os, Timur Tscherbul, and Micha\l{} Tomza for initial discussions.
\end{acknowledgments}

\bibliography{ca2f2}

\end{document}